\documentclass[prl,aps,twocolumn,epsf,psfig]{revtex4}
\usepackage[dvips]{epsfig}

\begin{document}

\title{Banded spatiotemporal chaos in sheared nematogenic fluids}

\author{Debarshini Chakraborty, Chandan Dasgupta, A.K. Sood}

\affiliation{Department of Physics, Indian Institute
of Science, Bangalore 560~012, India \\}


\begin{abstract}
We present the results of a numerical study of a model of the 
hydrodynamics of a sheared nematogenic 
fluid, taking into account the effects of order 
parameter stresses on the velocity profile, but allowing spatial 
variations only in the gradient direction. When parameter values are such 
that the stress from orientational distortions is comparable to the 
bare viscous stress, the system exhibits steady states 
with the characteristics of shear banding. In addition, 
nonlinearity in the coupling of extensional flow to orientation  
leads to the appearance of a new steady state in which the features of both 
spatiotemporal chaos and shear banding are present. 
\end{abstract}

\maketitle

\newpage

Experimental observations~\cite{expt1,expt2,expt3,expt4,expt5} of complex dynamics, including spatiotemporal 
chaos, in sheared wormlike micellar solutions have stimulated the development of several theoretical models.
The linearly extended nature of wormlike micelles leads naturally to considerations of models~\cite{th0,th1,us}
in which a nematic order parameter field is coupled to the hydrodynamic velocity.
Spatiotemporal rheochaos was demonstrated \cite{us} in the equations of passively 
sheared nematic hydrodynamics, with spatial variations allowed only in the gradient direction. This study, however,
did not find any clear signature of the formation of shear bands observed in 
experiments~\cite{expt2,expt3,expt4,expt5}. A different 
approach~\cite{smf1,smf2,smf3,smf4}, based on the Johnson-Segalman model~\cite{js}, with an added diffusive term (DJS model),
shows shear banding but no instability of the shear-banded state for spatial variations only in the gradient
direction. If spatial variations in the flow \cite{smf2} or vorticity \cite{smf3,smf4} directions are allowed, the shear-banded state
exhibits an instability that leads to complex, possibly chaotic dynamics.     
 
In this Letter, we present the results of a numerical study of a model~\cite{us} of the hydrodynamics
of a sheared nematogenic fluid in which spatial variations are allowed only in the gradient direction but
the assumption of passive advection~\cite{us} is removed, 
so that the effects of order parameter 
stresses on the velocity profile are fully taken into account. 
For parameter values such that the stress from orientational distortions is comparable to the bare viscous stress, 
the system exhibits a new steady state where we see the characteristics of shear banding 
in addition to the states seen earlier~\cite{us}.
Further, allowing nonlinearity in the coupling of extensional flow to orientation, 
leads to the appearance of new steady states. Among these new attractors of the
dynamics, the most significant one combines the features of both spatiotemporal chaos and shear 
banding. Thus, going beyond the passive advection approximation allows the occurrence of banded chaotic states
which were not observed in Ref.~\cite{us}.

In the simplifying limit where nonlinearities arising from the free-energy functional for nematic order are ignored, our model is equivalent to the DJS model studied in Refs.~\cite{smf1,smf2,smf3}, whose results we reproduce in the corresponding parameter range.
However, our fully nonlinear model 
also exhibits band formation in a different region of the parameter space and, in contrast to the results
reported in  Refs.~\cite{smf1,smf2,smf3}, instabilities of the banded state are found even when spatial
variations are allowed only in the gradient direction. These instabilities lead to a spatiotemporally 
chaotic state that also exhibits features of shear banding. Thus, order parameter nonlinearities
arising from the free-energy functional play a crucial role in our model, leading to the appearance
of new attractors with complex dynamics that are not present in the DJS model.


We now describe the model we consider and the numerical method used in our study. 
The nematic order parameter field $\textbf{Q(\textbf{r})}$ in our model is a traceless, symmetric 
second-rank tensor whose eigenvectors and eigenvalues
describe respectively the directions and magnitudes of local anisotropy.    
The equilibrium behavior of $\textbf{Q}$ is assumed to be governed by the
Landau-de Gennes free-energy functional  
\begin{eqnarray}
F[\textbf{Q}] &=& \int d {\bf r} \left[\frac{A}{2}\textbf{Q}:\textbf{Q} - 
\sqrt{\frac{2}{3}}B(\textbf{Q}\cdot\textbf{Q}):\textbf{Q} 
+\frac{C}{4}(\textbf{Q}:\textbf{Q})^2 \right .\nonumber\\ 
&+& \left . \frac{\Gamma_1}{2}\mbox{\boldmath $\nabla$} \textbf{Q}\ \vdots\ \mbox{\boldmath $\nabla$}\textbf{Q} 
+\frac{\Gamma_2}{2}\mbox{\boldmath $\nabla$}.\textbf{Q}.\mbox{\boldmath $\nabla$}.\textbf{Q} \right ]
\label{freeen}
\end{eqnarray}
with phenomenological parameters A, B and C governing the bulk free-energy difference between isotropic 
and nematic phases, and $\Gamma_{1}$ and $\Gamma_{2}$ related to the Frank elastic constants. 
In mean-field theory, the isotropic to nematic transition occurs when $A$ decreases below  $A_* = 2B^2/9C $.
The equation of motion obeyed by the alignment tensor is
\begin{equation}
\frac{\partial \textbf{Q}}{\partial t}+ \textbf{u}.\mbox{\boldmath $\nabla$}\textbf{Q} = 
\tau^{-1}\textbf{G} +(\alpha_0\mbox{\boldmath $\kappa$}+\alpha_1\mbox{\boldmath$\kappa$}.\textbf{Q})_{ST} 
-\mathbf{\Omega.Q}+\mathbf{Q.\Omega},
\label{eqnmotion}
\end{equation}
where the subscript $ST$ denotes symmetrization and trace removal, \textbf{u} is the hydrodynamic velocity field, 
\mbox{\boldmath $\kappa$} $\equiv (1/2)[\mbox{\boldmath {$\nabla$}}\textbf{u}+(\mbox{\boldmath{$\nabla$}}
\textbf{u})^T]$ and 
$\mathbf{\Omega} \equiv(1/2)$ $[\mbox{\boldmath {$\nabla$}} \textbf{u} -(\mbox{\boldmath{$\nabla$}}
\textbf{u})^T]$ are the deformation 
rate and vorticity tensors, respectively. The flow geometry imposed is plane Couette with velocity 
$ \mathbf{u} = \mathit {y\dot{\gamma}}\hat{x} $. We will refer to $\hat{\mathit{x}}$, $\hat{\mathit{y}}$ 
and $\hat{\mathit{z}}$ as the velocity ($\mathbf{u}$), gradient (\mbox{\boldmath$\nabla$}) and 
vorticity (\mbox{\boldmath $\omega$}) directions respectively. In Eq.(\ref{eqnmotion}), $\tau/A_*$ is a 
bare relaxation time and $\alpha_0$ and $\alpha_1$ are parameters related to flow alignment, originating 
in molecular shapes. 
Lastly, 
\begin{eqnarray}
\mathbf{G}&=& -\frac{\delta F}{\delta\mathbf{Q}}= -[A\textbf{Q}-\sqrt{6}B(\textbf{Q}.\textbf{Q})_{ST}
+C\textbf{QQ}:\textbf{Q}] \nonumber\\
&+& \Gamma_1\mbox{\boldmath $\nabla^2$}\textbf{Q} +\Gamma_2(\mbox{\boldmath $ \nabla\nabla$}.\textbf{Q})_{ST}
\label{geqn}
\end{eqnarray}
is the molecular field conjugate to \textbf{Q}. 
The contribution of the alignment tensor to the deviatoric stress
is given by
\begin{equation}
\mbox{\boldmath{$\sigma$}}^{\mbox{op}} = -\alpha_0 \textbf{G} - \alpha_1(\textbf{Q}.\textbf{G})_{ST}.
\label{stress}
\end{equation}
The total deviatoric stress is $\mbox{\boldmath{$\sigma$}}^{\mbox{op}}$ plus the bare viscous stress. 
Under the passive advection assumption, the bare viscous stress of the system is constant and it is sufficient 
to consider $\mbox{\boldmath{$\sigma$}}^{\mbox{op}}$ alone, as was done in Ref.~\cite{us}. 
To incorporate the full hydrodynamics in the problem, we now remove this assumption. 
We work in the Stokesian (zero Reynolds number) and incompressible limit, as is appropriate for 
reasonable experimental realizations of the systems of interest here. Thus,
\begin{equation}
\mathbf{\nabla}_j\mathbf{\sigma}^{\mbox{total}}_{ij}=0,
\label{stokes}
\end{equation}
where $\mbox{\boldmath{$\sigma$}}^{\mbox{total}}$ is the total stress tensor in the system, and 
\begin {equation}
\mbox{\boldmath{$\nabla$}}.\textbf{u}=0.
\label{incompress}
\end {equation} 
We consider spatial variation only along the $y$-axis. 
Then, Eq.(\ref{stokes}) with gradient terms involving derivatives only along the $y$-direction reduces to 
\begin{equation}
\mu\frac{\partial^2 {{u}_i}}{\partial y^2}= -\frac{\partial \mathbf{\sigma}^{\mbox{op}}_{yi}}{\partial y},
\label{stokes1}
\end{equation}
where $ i= x,z $, $\mu$ is the shear viscosity and
$ \mathbf{u}= y\dot{\gamma}\mathbf{\hat{x}} + \delta_1\mathbf{\hat{x}}+\delta_2\mathbf{\hat{z}} $, 
where $\delta_1$ and $\delta_2$ are $y$-dependent perturbations in the velocity profile. 
Since the fluid is incompressible, and spatial variation is only along the gradient axis, perturbations in 
${u_y}$ are zero. 

Following \cite{th1,us}, time is
rescaled by $\tau/A_* $ and $\textbf{Q}$ by $Q_k$, its magnitude at the transition temperature. 
Distances are rescaled by the diffusion length constructed out of $\Gamma_1$ and $A_*$. The ratio 
$\Gamma_2 / \Gamma_1$ is a parameter set to unity. We define a dimensionless viscosity parameter
$\eta \equiv \mu/(\alpha_0 \tau Q_k)$ and consider two cases, $\eta=1$ and $\eta=100$. The ratio of
the bare viscous stress to the stress from orientational distortions is determined by the quantity
$\eta \dot{\gamma}$ where $\dot{\gamma}$ is the dimensionless shear rate.

When the right-hand side of Eq.(\ref{stress}), including nonlinearities arising from 
the free-energy functional, Eq.(\ref{freeen}) and the $\alpha_1$ term, but excluding the gradient terms,
is linearized in the deviation
of \textbf{Q} from its uniform average value -- zero, if the underlying equilibrium phase is
isotropic --  
the derivatives of \textbf{Q} in the left-hand side 
of Eq.(\ref{eqnmotion}) are easily re-expressed as derivatives of the order-parameter stress. This equation, 
with further linearization of the nonlinear terms in $\textbf{G}$ and a redefinition of parameters, 
then becomes equivalent to the
constitutive equation for the viscoelastic stress in the DJS model. Thus, the DJS model may be thought of as a 
simplified (linearized)  version of the isotropic-phase limit of the model considered here
and the effect of the additional nonlinearities present in our model on the behaviour observed
in Refs.~\cite{smf1,smf2,smf3,smf4} becomes a question of 
considerable importance.

In our numerical study, a spatially discretized version of Eq.(\ref{eqnmotion}) is integrated forward in
time using a fourth-order Runge-Kutta algorithm. 
For much of this study 
we work at $A=0$ and $\alpha_1=0$, as in \cite{us}, so that the system in the absence of shear is deep in the 
nematic phase, in fact at the limit of metastability of the isotropic 
phase~\cite{footnote}. 
Our control parameters are $\lambda_k\equiv \sqrt{\frac{2}{3}}\alpha_0 $ related to the tumbling coefficient 
in Leslie-Ericksen theory \cite{th1}, and the dimensionless shear rate $\dot{\gamma}$. 
The two additional equations, Eq.(\ref{stokes1}), that enforce the Stokes condition, 
are solved simultaneously with the equation of 
motion, where the matrices, \mbox{\boldmath $\kappa$} and $\mathbf{\Omega}$ 
have terms involving $\delta_1$ and $\delta_2$. 
Starting from specified initial conditions, we let the equation of motion evolve and obtain the order parameter 
stress. Using $\mbox{\boldmath{$\sigma$}}^{\mbox{op}}$, Eq.(\ref{stokes1}) is then solved to construct the updated velocity profile, which is 
then fed back into the equation of motion.  Spatial derivatives are approximated by symmetric finite differences 
defined on the sites of a uniform mesh of $N$ points, 
Boundary conditions are so fixed that $\delta_1,\delta_2=0$ at the walls. Also, for defining derivatives 
at the mesh points $i=2$ and $i=N-1$, we set $f_0=f_1,f_{N+1}=f_N$, where $f$ is any variable of interest. 

We find that for $\eta=100$, this model reproduces the various steady states (``phases'') seen in 
\cite{us} with only small shifts in the phase boundaries. This is not surprising: for large values of $\eta$, 
the velocity corrections are small as can be seen from Eq.(\ref{stokes1}). 
For $\eta=1$, where order parameter stresses are comparable to those of the solvent, 
and $\alpha_1=0$, we find a new phase, 
as can be seen in Figs. 1 and 2. In this new phase, the steady state 
is a high-stress band with spatiotemporal or only temporal periodicity bounded by low-stress 
bands showing temporal periodicity (see Fig.2(e)). 
This new phase is a well-defined, banded periodic state between 
the irregular chaotic phase and the flow-aligned fixed point. The position of the high-stress band thus 
formed depends on the boundary conditions. The above case is seen when the order parameter 
tensor at the two ends of the system is aligned in the shear plane, irrespective of orientation. 
If, however, the tensor at the two ends is aligned parallel to the walls of the Couette cell, i.e. 
along $\hat{z}$, then a low-stress band with temporal periodicity is formed between two high-stress bands 
with either spatiotemporal or temporal periodicity near the walls. Space-time plots of the shear stress in
the phases found for $\eta=1$, $\alpha_1=0$ are shown in Fig. 2.  

 
\begin{figure} [!h]
\epsfig{file=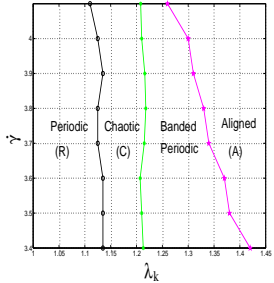,height=4cm,width=4cm,clip=} 
\caption{Phase diagram for $\eta=1.0$ and $\alpha_{1}=0$. Phase boundaries between regular periodic, chaotic,
banded periodic, and aligned (fixed point) attractors in the $\dot{\gamma}-\lambda_k$ plane are shown.}
\end{figure}
 
\begin{figure} [!h]
\begin{center}
\epsfig{file=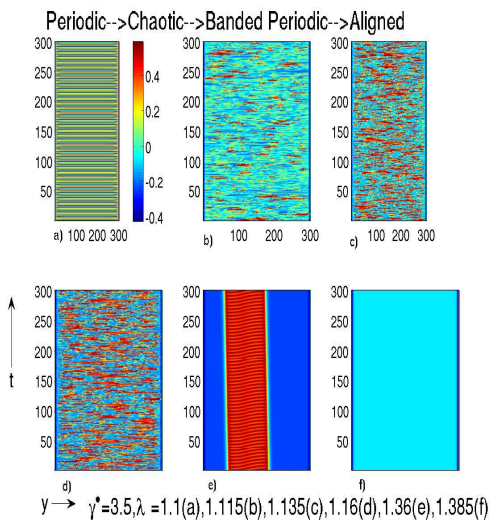,height=7cm,width=7cm,clip=}
\end{center}
\caption{Space-time plots of the shear stress in the phases observed [(a): regular periodic,
(b)-(d): chaotic, (e): banded periodic, (f): aligned] for a fixed shear rate, 
$\dot{\gamma}=3.5$ and varying $\lambda_k$, with $\eta=1,\alpha_{1}=0,A=0$.}
\end{figure}

For non-zero values of $\alpha_{1}$ and $\eta=1$, we obtain not only spatiotemporally periodic 
but also spatiotemporally chaotic attractors, where a high-stress band coexists with low-stress regions, 
as seen in Figs. 3(b) and 3(c) for $\alpha_1=0.3$. 
The position of the high-stress band is not fixed, and the steady state shows such a band of 
considerable width along the $y$-axis (gradient direction), meandering between the two walls of the 
Couette cell as a function 
of time. This spatiotemporally chaotic state (see below) with a wide high-stress band is found only for 
relatively low shear rates. At high shear 
rates ($\dot{\gamma} >3.7$), in the parameter-space region where a coexistence of high and low shear stress 
bands is first seen, 
the points of interfaces between the bands show complex oscillations in time. 
But as one moves deeper into this banded phase with 
increasing $\lambda_{k}$, the complex oscillations die down to simple periodicity and the order parameter stress 
exhibits only regular periodic character in the banded phase [Fig. 3(i)]. With further increase in 
$\lambda_k$, for a constant shear rate, the interfaces between the high-stress band and the adjoining lower 
stress bands merge and we see only temporally periodic oscillations, with a phase difference 
between the temporal oscillations at the merging point of the interface [Fig. 3(j)]. As 
$\lambda_k$ is increased further  with $\dot{\gamma}$ kept constant, chaotic oscillations build up from 
this temporally periodic state. Obtaining a complete phase diagram for $\alpha_1 \ne 0$ is
difficult, mainly because we find multiple locally stable attractors (and consequent dependence of the 
steady state on initial conditions) in some regions of the $\dot{\gamma}-\lambda_k$ plane.
\begin{figure}[!h]
\begin {center}
\epsfig{file=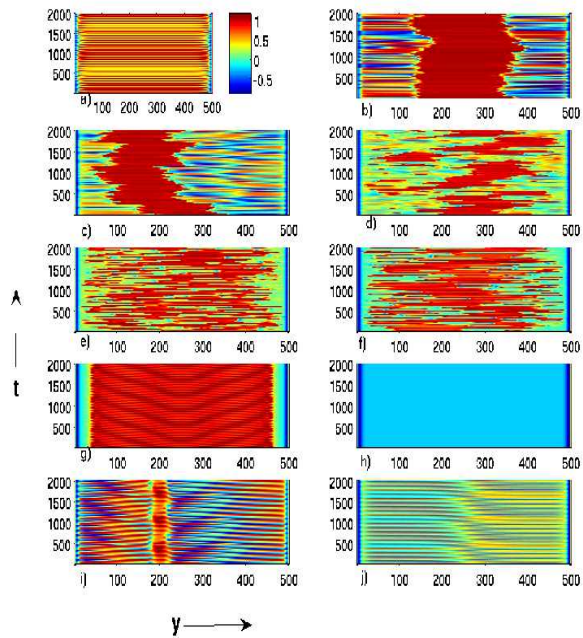,height=9cm,width=8.5cm,clip=}
\end{center}
\caption{Phases for $\dot{\gamma}=3.5$ with $\lambda_{k}=$ 0.888(a), 0.945(b), 1.025(c), 1.055(d), 1.07(e), 
1.137(f), 1.19(g), 1.33(h), $\eta=1.0,\alpha_{1}=0.3,A=0$. Space-time plots of the 
shear stress are shown in the
regular periodic [(a)], banded chaotic [(b) and (c)], chaotic [(d)-(f)], 
banded periodic [(g)] and aligned [(h)] phases. Phases for $\dot{\gamma}=4.1$ and $\lambda_{k}$ 
varying between 0.888 and 1.33 are the same as those for $\dot{\gamma}=3.5$, except for $\lambda_{k}$ between 
0.91(i) and 1.04(j). For $\dot{\gamma}=4.1$, the banded chaotic attractor of panel (b) evolves into a 
banded periodic attractor of panel (i), followed by a simple temporally periodic attractor with a kink 
[panel (j)], and  
the chaotic attractor of panel (e) as $\lambda_k$ is increased.}

\end {figure}



To characterize the chaotic states found, we study their Lyapunov spectra (LS). For a discrete N dimensional 
dynamical system, the $N$ Lyapunov exponents $\lambda_i,i=1:N$, arranged in decreasing order, form the LS. The 
number $N_{\lambda_+}$ of positive Lyapunov exponents and $\Sigma_{\lambda_{+}}$, the sum of the positive 
Lyapunov exponents, are useful quantities that can be calculated from the LS. 
In particular, $\Sigma_{\lambda_{+}}$ 
provides an upper bound and often a good estimate for the Kolmogorov-Sinai entropy that
quantifies the mean 
rate of growth of uncertainty in a system subjected to small perturbations \cite{er85}. Both these quantities 
scale extensively with system size in spatiotemporally chaotic systems. Since our system is an extended one 
with a large number of degrees of freedom, computing the LS is difficult owing to the 
inordinately large computing 
time and memory space required. The LS of a subsystem governed by the same equations of motion, 
when suitably rescaled,
can lead to the LS of the whole system. So instead of analyzing the whole system, we consider subsystems 
of comparatively small size $N_S$, at space points \textit{j} in an interval $i_0 \leq j \leq i_0+N_S-1$, 
where $i_0$ is an arbitrary reference point, and study the scaling of the relevant 
quantities with subsystem size $N_S$. For spatiotemporal chaos, it is also expected that with increasing 
subsystem size $N_S$, the largest Lyapunov exponent would 
increase, asymptotically approaching its value  
for subsystem sizes of the order of the system size. 
Our analysis shows that the embedding dimension at certain reference points can be 
so high that the scaling with subsystem size 
can only be partially studied due to computational constraints. However, 
over the limited range of subsystem sizes that we can access, we find,
depending on the choice of the reference point $i_0$, clear evidence for spatiotemporal 
chaos of varying dimensions in the banded state shown in Fig.3, panels (b) and (c). As shown in 
the top two panels of Fig.4, both $N_{\lambda_+}$ and 
$\Sigma_{\lambda_{+}}$ exhibit an approximately linear increase with 
increasing $N_S$ for the three reference points chosen. Also, the largest Lyapunov exponent (bottom panel of
Fig.4) shows the expected behavior as a function of subsystem size. 

 \begin{figure}[!h]
\begin{center}
\epsfig{file=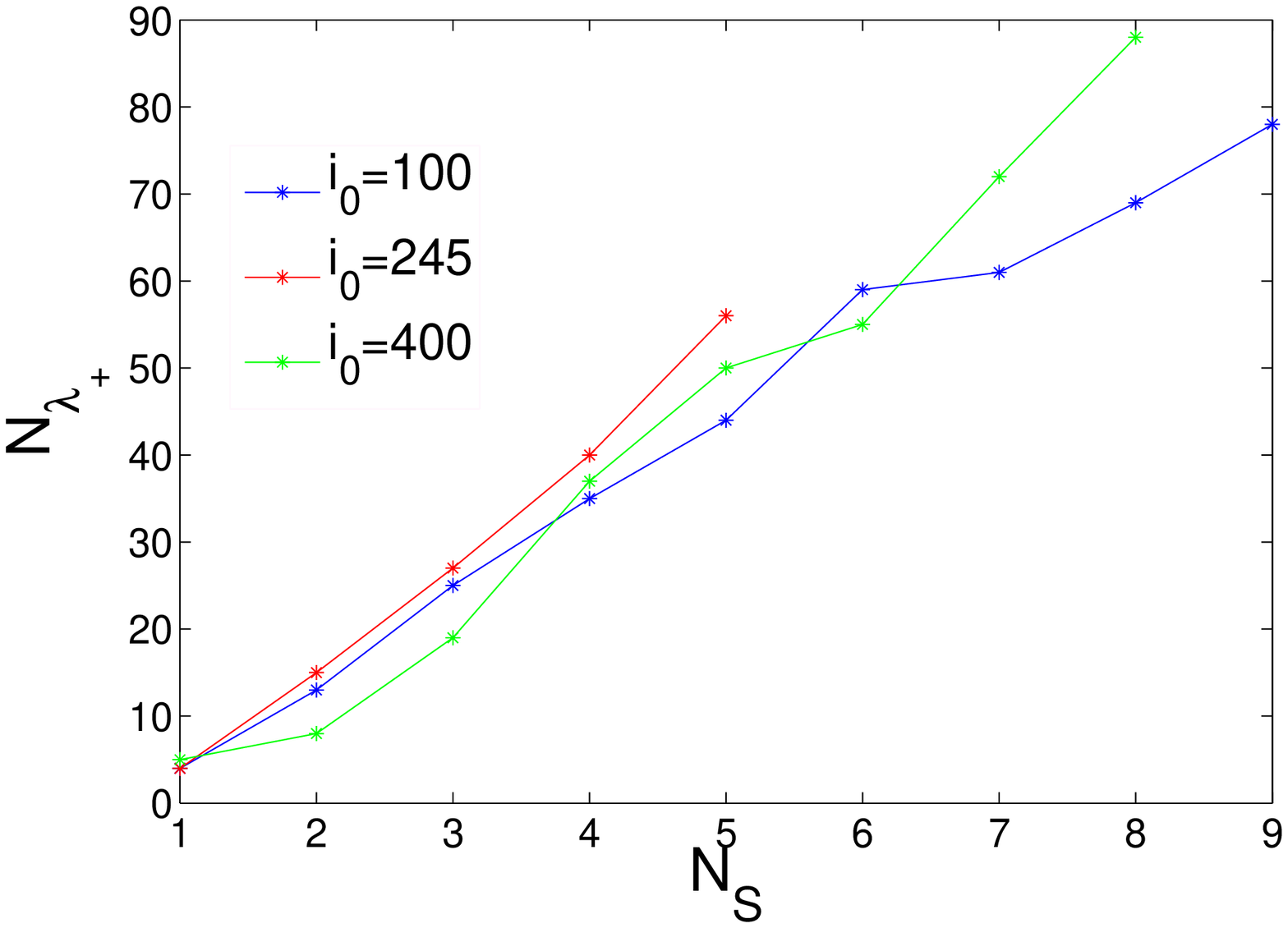,height=4cm,width=3.5cm,clip=}
\epsfig{file=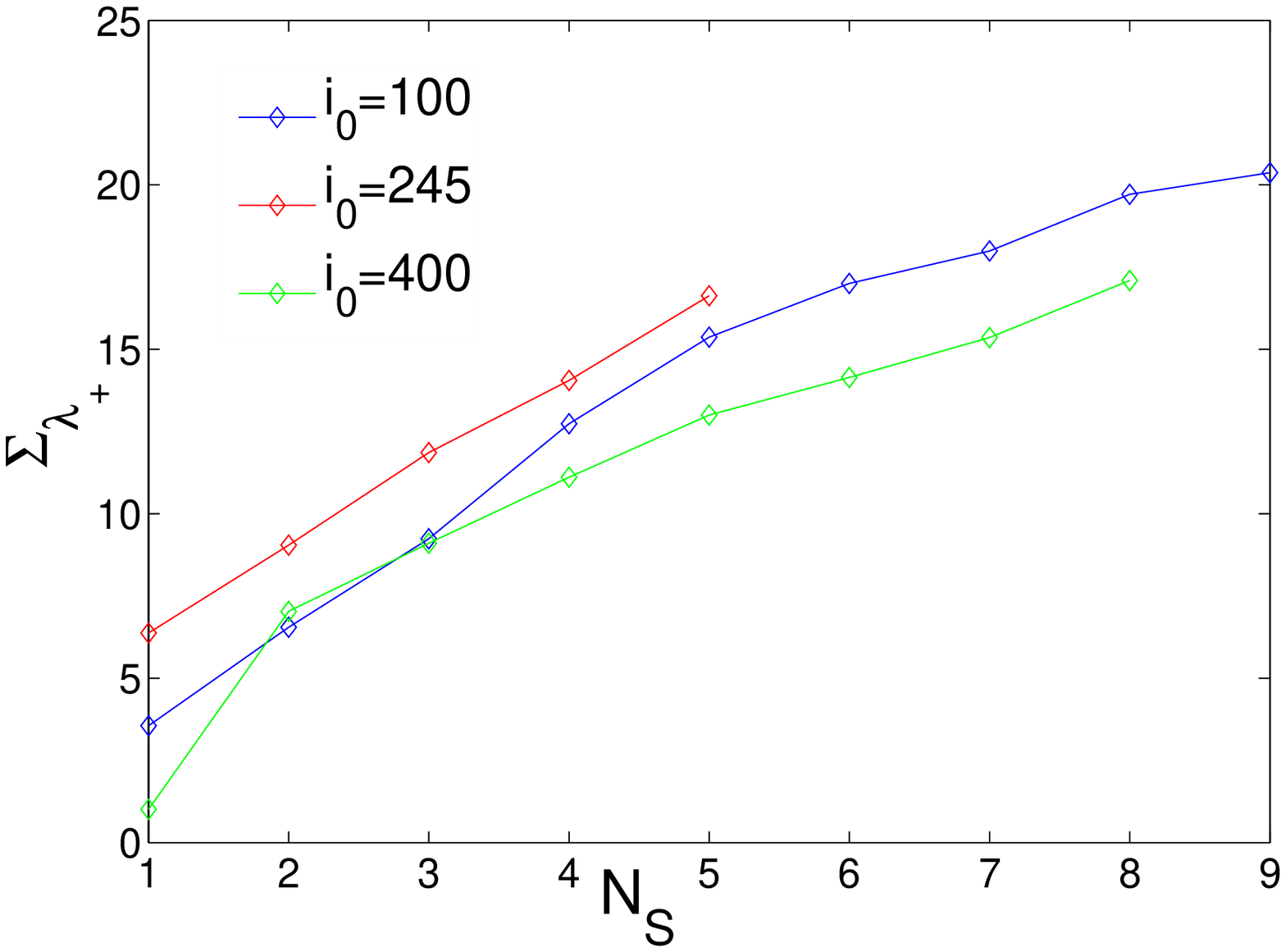,height=4cm,width=3.5cm,clip=}
\epsfig{file=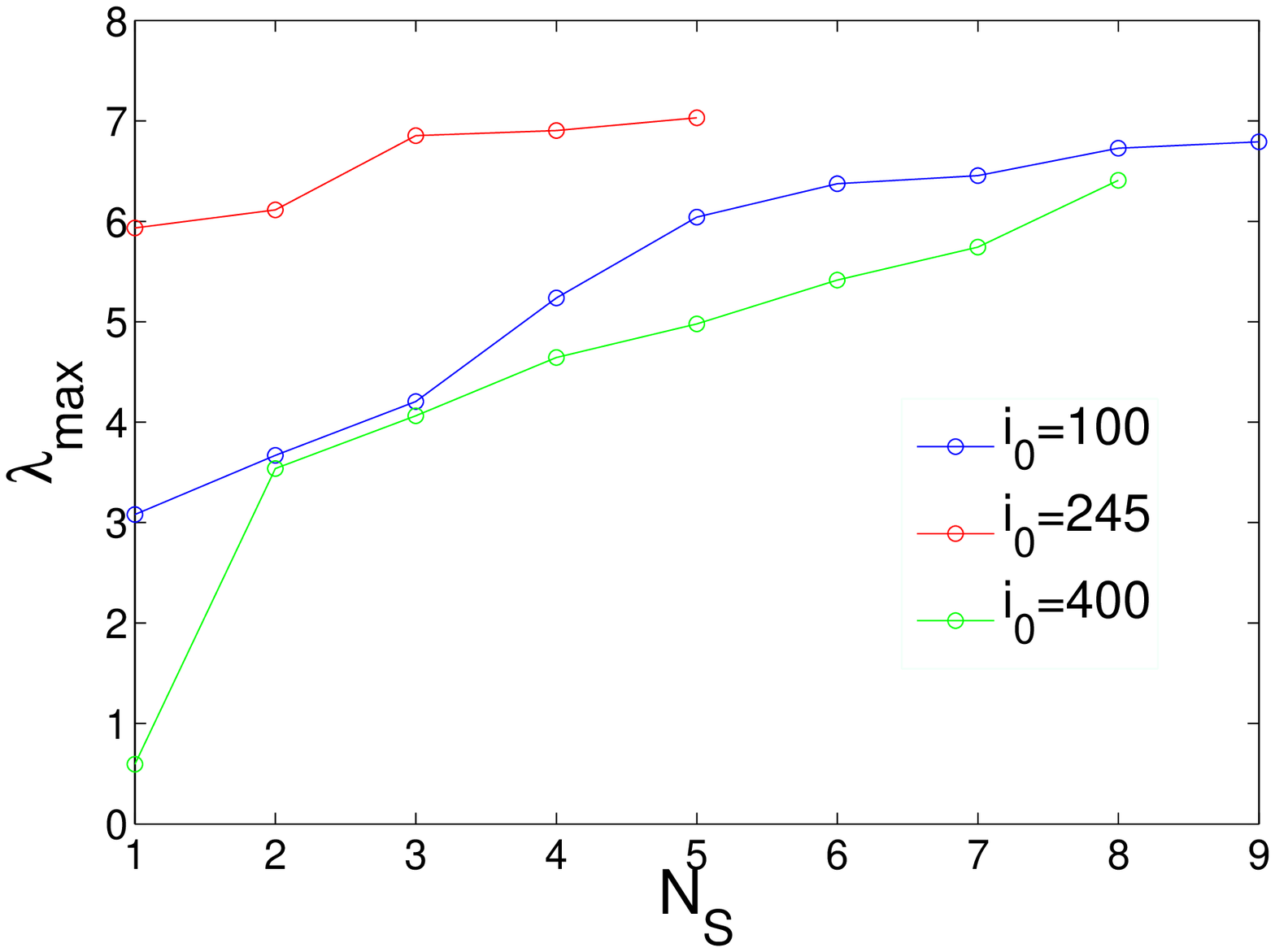,height=4cm,width=3.5cm,clip=}

 \end{center}
\caption{Number of positive Lyapunov exponents (top left panel), sum of the positive Lyapunov exponents, (top
right  panel), and the largest Lyapunov exponent (bottom panel) as functions of subsystem size $N_s$ for 
$\eta=1.0, \alpha_1=0.3,\dot{\gamma}=3.5,
\lambda_k=0.95$. Embedding dimension for the time-series at each space point is different for 
different $i_0$. The largest embedding dimension, 22, is for $i_0=245$, while those for $i_0=100, 400$ 
are 18 and 20 respectively.}
\end {figure}

As mentioned above, a linearized version of our model is equivalent to the DJS model considered by Fielding 
and co-workers~\cite{smf1,smf2,smf3,smf4}, and we have reproduced their main results for a similar choice of
parameters. In particular, we have found that the shear-banded state found by them remains stable when the 
additional nonlinearities in our model are included. However, the values of $\eta$ for which we find the
banded attractors in Figs. 2 and 3 lie outside the range in which the linearized model exhibits shear banding.
Also, in contrast to the results of Refs.~\cite{smf1,smf2,smf3,smf4}, our nonlinear model exhibits an instability of
the banded state as $\lambda_k$ is increased, even if spatial variations are allowed only in the gradient
direction (see Figs. 2 and 3). These results imply that the physics of the formation of the shear banded state,
its instability, and the coexistence of shear banding and spatiotemporal chaos (for $\alpha_1 \ne 0$) 
in our nonlinear model is very
different from that of the DJS model. Thus, our order-parameter based model, which includes the DJS model as
a special case, exhibits additional complex collective dynamics arising from nonlinearities in
the free energy that describes the equilibrium behavior of the order parameter. Development of similar models 
for rheological chaos in other complex fluids would be most interesting. It would also be worthwhile to explore
the consequences of allowing spatial variations in the flow and vorticity directions in our model.  

We are grateful to Sriram Ramaswamy for many helpful discussions.


\begin{thebibliography}{99}
\bibitem{expt1}R. Bandyopadhyay, G. Basappa, and A. K. Sood, 
Phys. Rev. Lett. {\bf 84}, 2022 (2000); 
R. Bandyopadhyay and A. K. Sood, Europhys. Lett. {\bf 56}, 447 (2001).
\bibitem{expt2} R. Ganapathy and A. K. Sood, Phys. Rev. Lett. {\bf 96}, 108301 (2006); 
R. Ganapathy, S. Majumdar and A. K. Sood, Phys. Rev. E {\bf 78} 021504 (2008).  
\bibitem{expt3} L. Becu, S. Manneville, and A. Colin, Phys. Rev. Lett. {\bf 93}, 018301 (2004).
\bibitem{expt4} M.R. Lopez-Gonzales {\it et al.}, Phys. Rev. Lett. {\bf 93}, 268302 (2004); S. Lerouge, M. 
Argentina, and J. P. Decruppe, Phys. Rev. Lett. {\bf 96}, 088301 (2006); M.A. Fardin {\it et al.}, Phys. Rev.
Lett. {\bf 103}, 028302 (2009).
\bibitem{expt5}P. Nghe, S. M. Fielding, P. Tabeling, A. Ajdari, arXiv:0909.1306v1, (2009).
\bibitem{th0}M. Grosso, R. Keunings, S. Crescitelli, and P. L. Maffetone, Phys. Rev. Lett. {\bf 86}, 3184 (2001). 
\bibitem{th1}G. Rien\"acker, M. Kro\"ger, and S. Hess, Phys. Rev. E {\bf 66}, 040702(R) (2002); Physica A {\bf 315}, 537 (2002). 
\bibitem{us} M. Das, B. Chakrabarti, C. Dasgupta, S. Ramaswamy and 
A.K. Sood, Phys. Rev. Lett. {\bf 92},055501 (2004); Phys. Rev. E {\bf 71},021707 (2005).
\bibitem{smf1}S.M. Fielding, Phys. Rev. Lett. {\bf 95}. 134501 (2005).
\bibitem{smf2}S.M. Fielding and P.D. Olmsted, Phys. Rev. Lett. {\bf 96}, 104502 (2006)
\bibitem{smf3}S.M. Fielding, Phys. Rev. E {\bf 76}, 016311 (2007).
\bibitem{smf4}S.M. Fielding,  arXiv:0912.2322v1, (2009). 
\bibitem{js} M. W. Johnson and D. Segalman, J. Non-Newtonian Fluid Mech. {\bf 43} 311 (1977)  
\bibitem{footnote} We have found qualitatively similar behavior for $A$ slightly larger than $A_*$,
for which the equilibrium phase in the absence of shear is isotropic.
\bibitem{er85}J.P. Eckmann and D. Ruelle, Rev. Mod. Phys. {\bf 57}, 617 (1985) 


\end{thebibliography}
\end{document}